\documentclass[pdflatex,sn-mathphys-ay]{sn-jnl}%

\usepackage{graphicx}%
\usepackage{multirow}%
\usepackage{amsmath,amssymb,amsfonts}%
\usepackage{amsthm}%
\usepackage{mathrsfs}%
\usepackage[title]{appendix}%
\usepackage{xcolor}%
\usepackage{textcomp}%
\usepackage{manyfoot}%
\usepackage{booktabs}%
\usepackage{tabularx}%

\graphicspath{{../figures/}}

\newcommand{\rcr}{\ensuremath{R_{\mathrm{CR}}}}
\newcommand{\rtw}{\ensuremath{R_{25}}}

\newcommand{\msq}{\ensuremath{m=2}}

\theoremstyle{thmstyleone}%

\theoremstyle{thmstyletwo}%

\theoremstyle{thmstylethree}%

\begin{document}

\title[Automated photometric corotation radii in spiral galaxies]%
{An automated photometric pipeline for measuring the corotation radius of spiral galaxies}

\author*{Kimia Yazdani}
\author{Sahel Mesforoush}
\author{Wayne Hayes}


\abstract{The corotation radius ($R_{\rm CR}$) is a fundamental property of spiral galaxies, marking the location where the angular speed of the spiral pattern matches the orbital speed of the disk material. Despite its importance for understanding spiral structure and galaxy evolution, corotation measurements remain available for only a relatively small number of galaxies and are often derived using heterogeneous techniques. We present a fully automated, purely photometric pipeline for identifying one or more corotation radii by tracking the azimuthal phase offset between young and old stellar populations as a function of galactocentric radius. Building on the color-gradient framework of \citet{Puerari1997}, we extract azimuthal brightness profiles from deprojected dual-band images, fit periodic models to the profiles, and identify corotation candidates at radii where the phase difference between the two stellar tracers changes sign. Each candidate is assigned an adaptive uncertainty interval and reported with a goodness-of-fit metric. We benchmark the method against a cross-method literature compilation, scoring every technique -- including our own -- against a reference built from the published measurements alone and evaluated leave-one-out. Over the $90$ galaxies for which such a reference exists, the pipeline places a correct corotation radius in $80.0\%$ of them, recovers $63.6\%$ of their known corotation radii, and reaches a harmonic mean of recall and purity of $0.58$, the highest values of any method in the comparison. Unlike the published methods, each of which was applied to a small, curated set of galaxies, our pipeline is run on the entire reference sample without any pre-selection, a sample two to six times larger than existing methods. The pipeline is scalable to large surveys, flexible with respect to spiral mode, and naturally recovers multiple corotation radii in galaxies with complex spiral structure, providing a uniform framework for studying disk resonances across diverse galaxy populations.}


\maketitle


\section{Introduction}\label{sec:intro}

The origin and persistence of spiral structure in disk galaxies remain
unresolved problems in galactic dynamics. Two main frameworks are
currently discussed in the literature. The first is the
quasi-stationary density wave (QSDW) theory proposed by \citet{Lin1964}, in which spiral arms are interpreted as density waves rotating
with a constant pattern speed, $\Omega_{\mathrm{p}}$, independent of
galactocentric radius. In this picture, the corotation radius (CR)
marks the location where the angular velocity of the disk material
equals the pattern speed, dividing the disk into two dynamical
regions: inside CR, the material overtakes the pattern, while outside,
the pattern moves faster than the material.

The alternative view is that spiral arms are transient and recurrent
features \citep{Sellwood2011,Baba2013}. In this model, arms
continuously form, shear, and dissipate under the influence of
differential rotation and local gravitational instabilities, leading
to a pattern speed that varies with radius.
Recent simulations,
together with kinematic analyses of observed galaxies \citep{Meidt2009}, and the diversity of observed spiral
morphologies, suggest that multiple mechanisms are at play in shaping
spiral structure \citep{Dobbs2014}.
Bar-driven or tidally induced
spirals may coexist with dynamically formed transient arms within the
same disk, indicating that the formation and maintenance of spiral
patterns cannot be explained by a single universal process.

The corotation radius plays a critical role in discriminating between
these models and in understanding the evolution of galaxies. The CR is
a key site of angular momentum exchange between the disk and the
spiral pattern \citep{Sellwood2002} and can influence stellar
migration, chemical gradients, and orbital stability \citep{Contopoulos2013,Scarano2013}. Its location relative to the
bar also provides constraints on the dynamical coupling between the
stellar disk and the dark matter halo \citep{Debattista2000}.

Despite extensive theoretical and observational efforts, reliable
determination of the corotation radius remains challenging. A variety
of techniques have been developed, including the
Tremaine--Weinberg method \citep{Tremaine1984}, the
potential--density phase-shift approach \citep{Zhang2007}, and
photometric diagnostics based on color gradients or age offsets
\citep{Puerari1997,Tamburro2008}. However, these
methods often produce inconsistent results, reflecting
observational limitations
\citep{Vallee2020,Williams2021}.

Recent meta-analyses of published corotation measurements further
emphasize the complexity of this problem. A cross-method survey of
$547$ galaxies compiling $1{,}711$ published measurements \citep{Kostiuk2024} revealed that
different methods agree on a consistent corotation radius for only
about $15$--$25\%$ of galaxies, while for the majority the measurements
from different techniques are widely scattered or produce multi-peaked
distributions of \rcr\ estimates. In roughly $10\%$ of the sample,
multiple localized corotation radii were found, consistent with the
presence of several spiral modes rotating at distinct pattern speeds.
These results provide compelling observational support for the
presence of multiple corotation resonances within individual galaxies,
consistent with either the coexistence of multiple driving mechanisms
or a single mechanism --- such as recurrent, transient spirals --- that
produces multiple pattern speeds over time. We use the per-galaxy measurements from this compilation as the reference set for validating our method (\S \ref{sec:results}), restricted to the $161$ galaxies with corotation measurements from at least two different methods, since a cross-method comparison requires more than one method per galaxy.

The increasing evidence for complex, multi-mode spiral dynamics
highlights the need for automated and scalable techniques capable of
systematically identifying multiple corotation radii across large
samples. To address this, we present a new, purely photometric pipeline
that quantifies the asymmetry of the surface-brightness profile across
spiral arms, an effect that reverses at the corotation radius due to
the change in the relative motion of stars and gas. By measuring this
asymmetry, the method can locate one or more corotation radii without
requiring kinematic data, making it suitable for both nearby and
distant galaxies. The pipeline is fully automated and optimized for
large survey datasets, enabling consistent detection of multiple
resonance positions and facilitating statistical analyses of
corotation radii in relation to galaxy morphology and dynamics. We
validate the method using a set of local galaxies with existing
corotation measurements and demonstrate its scalability for
application to large samples.

Our method follows the conceptual framework of \citet{Puerari1997}, which identifies the corotation radius through the reversal of color gradients across spiral arms. In their picture, the relative azimuthal phase between young (blue) and old (red) stellar populations changes sign at the location where the angular velocity of the spiral pattern matches that of the disk material. This provides a purely photometric route to corotation, independent of kinematic data. We measure color gradients directly; stellar age is not measured but inferred from the well-established color--age relation, in which younger stellar populations are bluer and older populations are redder. While the original \citeauthor{Puerari1997} method demonstrated this principle for a small number of manually analyzed galaxies, its scalability was limited by image noise, subjective phase tracing, and fixed model assumptions.

In this study, we generalize and automate the \citeauthor{Puerari1997} approach to enable systematic application across large galaxy samples. Our pipeline extracts azimuthal brightness profiles in two wavebands, fits the underlying periodic structure with a parametric function, and computes the radial phase difference $\Delta\phi(R)$ between the bands to locate the radii at which it changes sign. Unlike previous implementations, both the spiral mode number ($m$) and the functional form of the fitting model are configurable.

The remainder of the paper is organized as follows.
Section~\ref{sec:methods} describes the photometric pipeline, from the
extraction of azimuthal brightness profiles to the identification and
validation of candidate corotation radii. Section~\ref{sec:results}
presents the validation of the method against a cross-method
literature compilation and quantifies its accuracy at the galaxy level
and per corotation radius (per-\rcr). Section~\ref{sec:discussion} discusses the
implications of these results for the dynamical interpretation of
spiral structure and outlines future extensions of the framework.


\section{Methods}\label{sec:methods}

The corotation radius (\rcr) of a spiral galaxy is an important
dynamical parameter. It marks the distance from the galactic center
where the angular velocity of the spiral pattern matches that of the
stars and gas orbiting within it. At this radius, the phase difference
between tracers of the spiral pattern, such as older (redder) stars,
and tracers of recent star formation, such as younger (bluer) stars,
is expected to reach zero and switch sign.

To identify \rcr, we developed an automated pipeline that measures
this phase difference as a function of galactocentric radius. The
overall procedure consists of five main steps: (1) extracting
azimuthal brightness profiles from images in two wavebands;
(2) modeling these profiles to estimate their phase;
(3) computing the phase difference between the two wavebands;
(4) identifying zero-crossings that correspond to candidate corotation
radii; and (5) assigning each candidate an uncertainty interval and a
goodness-of-fit measure. Each of these steps is described in detail below.

\subsection{Azimuthal brightness profile extraction}\label{sec:methods:profile}

We begin with deprojected (face-on) galaxy images. We work from the calibrated FITS images, which keep the full dynamic range of the data and a known pixel scale, so that radii can be placed on a physical scale. Both wavebands are resampled onto a common grid in sky coordinates centred on the galaxy, and deprojected with a single geometric transform derived from the second moments of the light distribution in the longer-wavelength image, which traces the smooth old stellar disk. Applying one transform to both bands guarantees that they are deprojected identically, so that any residual error in the deprojection cancels in the phase difference. The images are cutouts spanning $3\,\rtw$, so that the full radial range sampled below lies inside the image; rings that extend beyond the image footprint are discarded. For
our analysis, we primarily use SDSS $g$-band and $i$-band images,
which serve as proxies for young and older stellar populations,
respectively. For the galaxies that lie outside the SDSS
footprint we substitute the closest available equivalents from
all-sky surveys -- typically DSS2 blue for the young tracer and
WISE $3.4\,\mu$m for the old tracer -- so the wavelength
baseline is not identical across the sample. Because the measured
phase difference grows with the separation in wavelength between the
two tracers, results for the two subsamples are not directly
interchangeable.

\begin{figure}[t]
\centering
\includegraphics[width=\textwidth]{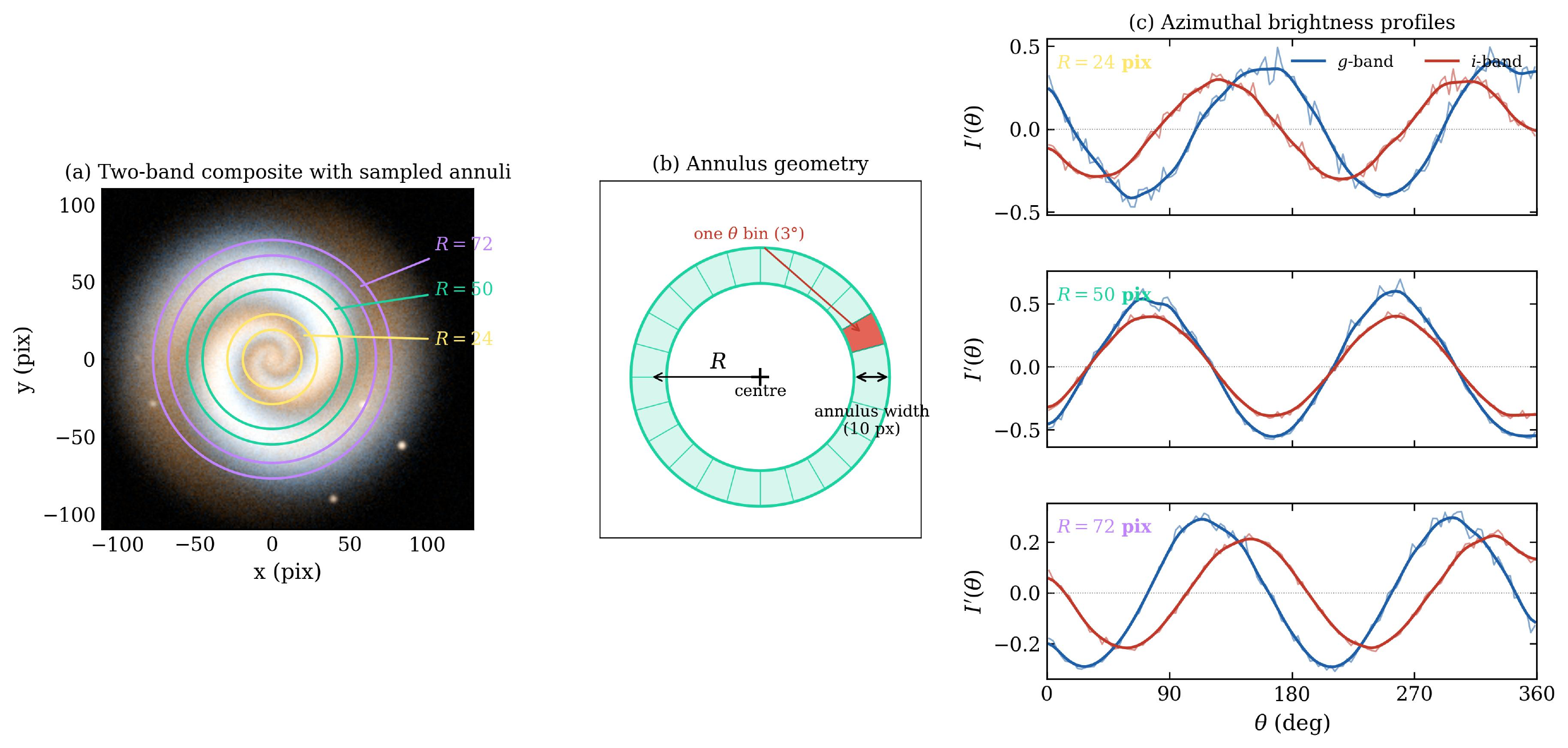}
\caption{Construction of the azimuthal brightness profiles.
\textbf{(a)} Two-band composite of a deprojected synthetic spiral
galaxy with three sampled annuli highlighted in color. \textbf{(b)}
Schematic of a single annulus, showing its radial width, set to $0.04\,\rtw$ in the analysis and drawn here schematically, and the partition into $120$ equal-azimuth bins of
$3^{\circ}$ each, one of which is highlighted. \textbf{(c)}
Mean-subtracted azimuthal brightness profiles $I_R'(\theta)$ in the
$g$- and $i$-bands at the three sample radii. The thin lines show the
raw mean-subtracted profiles, while the thick lines show
Gaussian-smoothed versions for display. Note that the blue curve shifts leftward from the top panel to the bottom (i.e., with increasing radius).}
\label{fig:methods:profile}
\end{figure}

For each galaxy, we analyze a set of discrete galactocentric radii,
$R$ (see Fig.~\ref{fig:methods:profile}). Radii are
chosen in units of each galaxy's own size: we take $34$ radii spaced
uniformly over $0.08\,\rtw < R < 1.25\,\rtw$, with an annulus width of
$0.04\,\rtw$. Sampling in units of \rtw\ ensures that the same radial grid corresponds to the same relative position in every disk. We take \rtw\ from the PGC catalogue \citep{Paturel2003}. At each radius $R$, we
compute the azimuthal profile as follows. The pixels within this annulus are divided into $120$
azimuthal bins, each spanning $3^{\circ}$. For each bin, we compute
the mean brightness, which gives the azimuthal brightness profile,
$I_R(\theta)$, for that waveband. We then normalize each profile by
subtracting the mean brightness of the annulus,
\begin{equation}
I_R'(\theta) = I_R(\theta) - \langle I_R \rangle_{\mathrm{annulus}},
\label{eq:normprof}
\end{equation}
so that $I_R'(\theta)$ encodes only the azimuthal modulation produced by
spiral structure. Repeating this process across all radii yields a set
of normalized azimuthal brightness profiles for both the blue
($g$-band) and red ($i$-band) images.

\subsection{Quantifying the spiral arm phase}\label{sec:methods:phase}

More flexible descriptions of the azimuthal profile, such as a Fourier
decomposition with several harmonics or a wavelet analysis, can in
principle capture asymmetric or multi-armed structure. In practice,
however, these models also fit the small-scale fluctuations in the
profile --- individual H\,\textsc{ii} regions, foreground stars, dust
lanes and photon noise --- and the azimuthal position of the arms read
from such a reconstruction is pulled by every component that is
included. In our tests this shifted the inferred position of the
two-armed pattern from radius to radius in a way that did not reflect
the spiral structure itself. We therefore model each profile with a
single $m=2$ cosine. Restricting the model in this way forces all of
the azimuthal structure, including the noise, to be described by one
mode, so that the fitted phase measures the position of the dominant
two-armed pattern alone, while the remaining structure is left in the
residuals rather than in the arm position.

To estimate the azimuthal position of the spiral
arms at each radius, we model the normalized brightness profile
$I_R'(\theta)$ with a cosine
function. This choice is motivated by the fact that spiral arms
introduce a periodic signal in azimuth, and a cosine function provides
a simple way to capture that structure while still allowing a
well-defined phase. We assume that the dominant structure is a
two-armed spiral pattern (\msq). We fit the binned profile
directly; smoothing it with a one-dimensional Gaussian filter before
fitting changes the results negligibly.

\begin{figure}[t]
\centering
\includegraphics[width=0.95\textwidth]{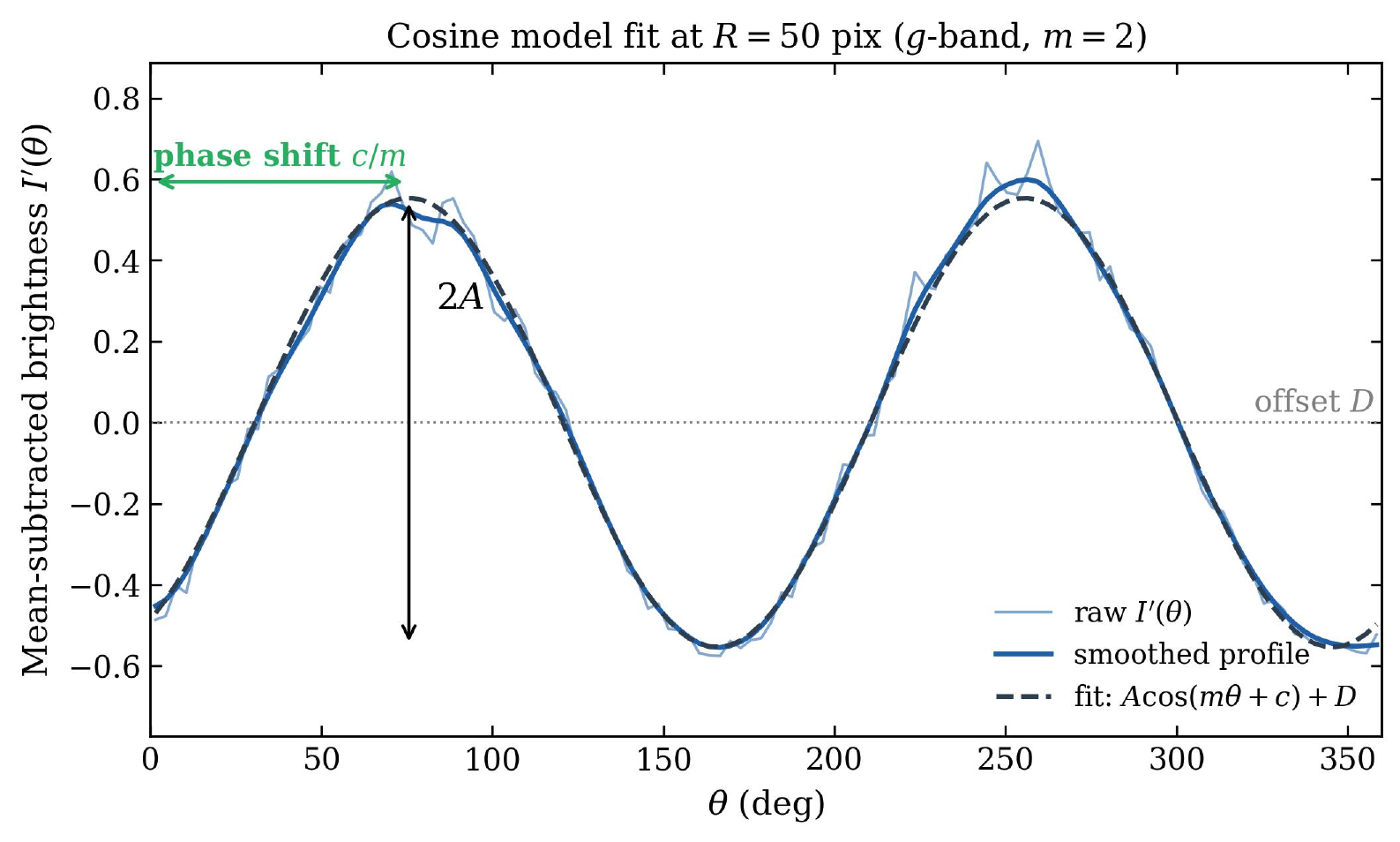}
\caption{Anatomy of the cosine fit at a single annulus. The light
curve shows the raw mean-subtracted brightness profile $I_R'(\theta)$,
the bold curve the smoothed version, and the dashed black curve the
best-fit cosine model Eq. \ref{eq:cosfit} with $m=2$. The peak-to-peak
amplitude $2A$, the constant offset $D$, and the phase shift $c/m$
that defines the angular position of the arms are indicated.}
\label{fig:methods:cosfit}
\end{figure}

We then fit the binned profile with the function
\begin{equation}
f(\theta) = A\,\cos\!\left(m\,\theta + c\right) + D,
\label{eq:cosfit}
\end{equation}
where $\theta$ is the azimuthal angle,
$A$ is the amplitude, $m$ is the mode number (fixed to $2$ in this
work), $c$ is the phase shift, and $D$ is a vertical offset (Fig.~\ref{fig:methods:cosfit}). While the
default form is a single cosine, both $m$ and the analytic form of the
fitting function are user-configurable parameters of the pipeline,
which allows the framework to be extended to higher-order or
asymmetric spiral patterns.

\begin{figure}[t]
\centering
\includegraphics[width=0.85\textwidth]{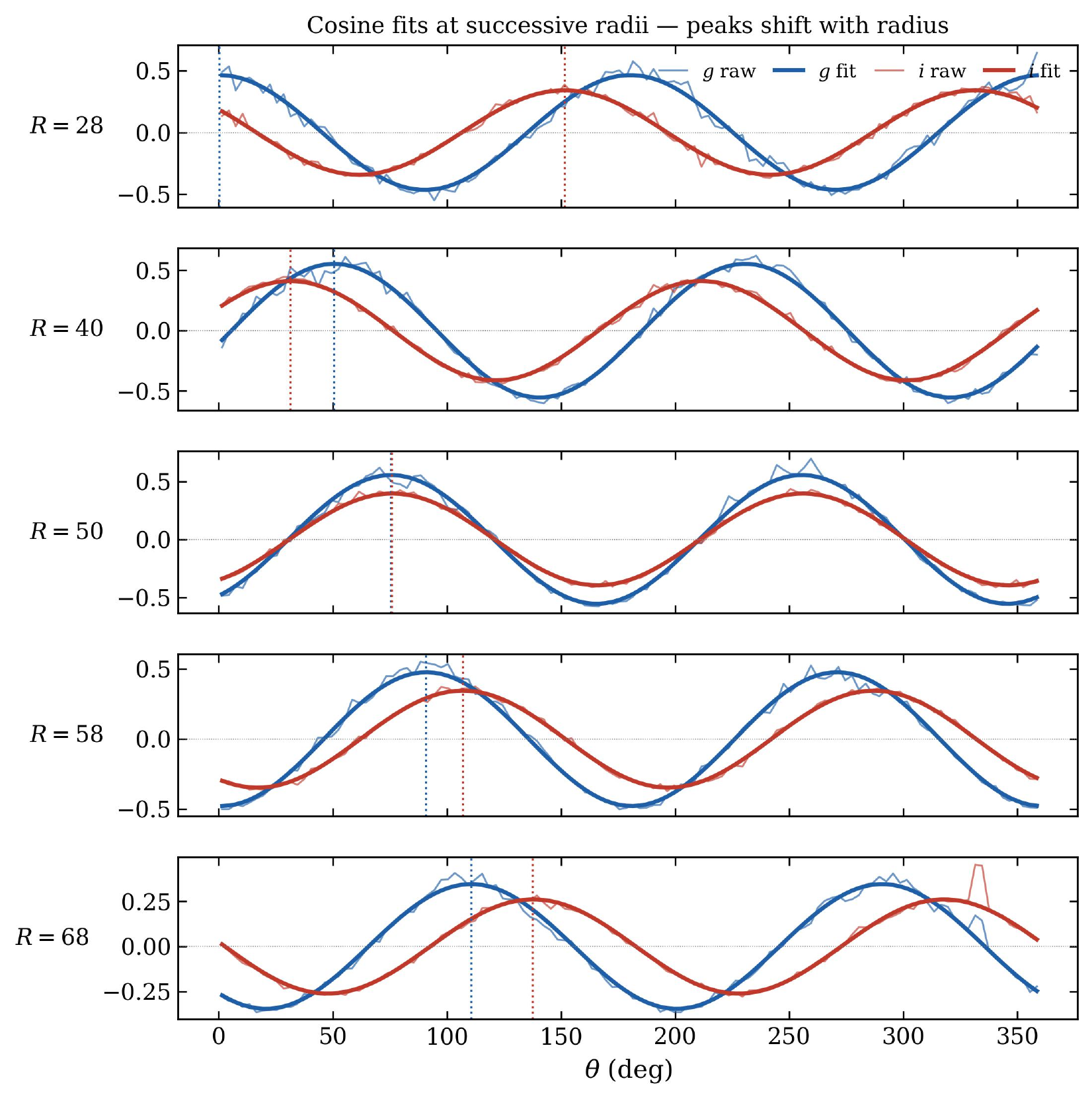}
\caption{Schematic evolution of the cosine fits across radii in both bands.
Each panel shows the raw azimuthal profiles (thin lines) and the
corresponding cosine fits (thick lines) for the $g$-band (blue) and
$i$-band (red) at increasing galactocentric radius. The dotted
vertical lines mark the angular position of the fitted peak in each
band; their relative shift with radius reflects the radial
dependence of the phase difference between the two stellar
populations.}
\label{fig:methods:cosevol}
\end{figure}

The main quantity we extract from the fit is the phase shift $c$. The
fitting procedure is applied independently to the blue and red
profiles at each radius $R$, yielding two functions
$c_{\mathrm{blue}}(R)$ and $c_{\mathrm{red}}(R)$ (Fig.~\ref{fig:methods:cosevol}). Each band is fitted from a starting point derived from its own profile.

Two conventions are required to make the fitted phase well defined.
First, the model is invariant under $A \rightarrow -A$, $c \rightarrow
c + \pi$, so we canonize every fit to $A > 0$, adding $\pi$ to $c$
where necessary; without this the two bands may be returned on
opposite branches and the inferred phase difference is displaced by
$\pi$. Second, the azimuthal position of an $m$-fold pattern is
defined only modulo $2\pi/m$, so the phase difference of
Eq.~\ref{eq:dphi} is wrapped into the interval
$(-\pi/m, +\pi/m]$ before any zero-crossing is sought.

Because the phase measurements depend on how well the cosine model
describes the brightness profile, we use the mean squared error (MSE)
of each fit as a quality metric. A high MSE indicates that the model
does not describe the brightness profile well, which can occur when
the spiral structure is weak, noisy, or not adequately represented by
a simple two-armed pattern. In such cases, the phase estimate is
considered unreliable.

Candidate corotation radii (Section~\ref{sec:methods:dphi}) located
near regions with high MSE are flagged as low confidence and reported
alongside the radius itself, so that users of the catalogue can apply
their own quality threshold.

Concretely, the MSE of each band is normalized by the mean brightness
of the corresponding annulus, and we define a per-radius confidence as
the reciprocal of the larger of the two normalized residuals. 

\subsection{Phase difference and corotation radius
identification}\label{sec:methods:dphi}

\begin{figure}[t]
\centering
\includegraphics[width=\textwidth]{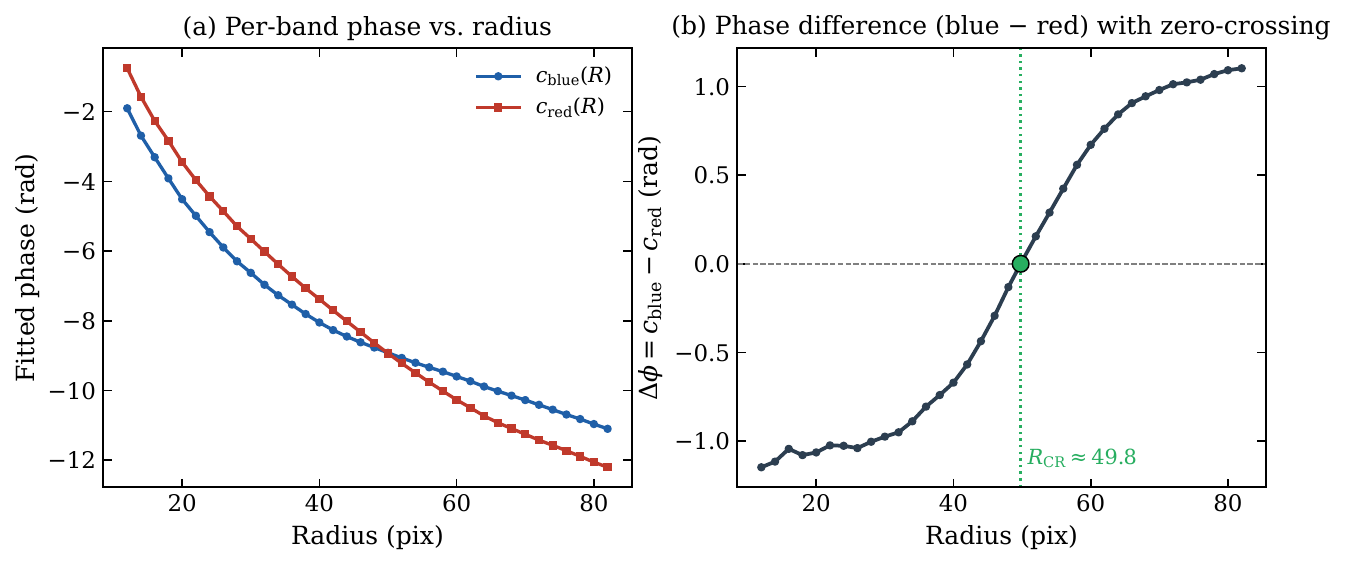}
\caption{Per-band phase as a function of radius and the resulting
phase difference. \textbf{(a)} Fitted phase shifts
$c_{\mathrm{blue}}(R)$ and $c_{\mathrm{red}}(R)$ obtained
from the $g$- and $i$-band fits. \textbf{(b)} Phase
difference $\Delta\phi(R)=c_{\mathrm{blue}}(R)-c_{\mathrm{red}}(R)$.
Zero-crossings of $\Delta\phi(R)$, marked by green points, define the
candidate corotation radii of the galaxy.}
\label{fig:methods:dphi}
\end{figure}

We define the phase difference between the young (blue) and old (red)
stellar populations as
\begin{equation}
\Delta\phi(R) = c_{\mathrm{blue}}(R) - c_{\mathrm{red}}(R).
\label{eq:dphi}
\end{equation}
This phase difference is the key quantity for identifying corotation.
The corotation radius \rcr\ is defined as the radius at which the
phase difference changes sign,
\begin{equation}
\Delta\phi(\rcr) = 0,
\label{eq:rcr}
\end{equation}
i.e.\ the radius at which the two populations are aligned in azimuth, with $\Delta\phi(R)$ having opposite signs on either side of \rcr.
A galaxy may possess one or several such radii, depending on the
complexity of its spiral structure (Fig.~\ref{fig:methods:dphi}).

\subsection{Automated zero-crossing detection and uncertainty
estimation}\label{sec:methods:uncertainty}

To locate the zero-crossings we scan $\Delta\phi(R)$ for changes of sign between adjacent sampled radii. Each such bracketing pair defines one candidate \rcr, whose position is obtained by linear interpolation between the two radii that bracket it; no smoothing or resampling onto a finer grid is performed. A candidate is recorded only where $\Delta\phi(R)$ changes sign between adjacent radii (or is exactly zero at a sampled radius); a curve that approaches zero and turns back without changing sign does not produce a candidate.

Candidates are further restricted to the radial range
$0.15\,\rtw < \rcr < 0.60\,\rtw$. This is a noise filter. Towards the
centre the annuli contain few independent resolution elements and are
dominated by the bulge, while in the outermost annuli the arm amplitude
falls towards the sky background; in both tails the fitted phase is
poorly constrained, so $\Delta\phi$ fluctuates and crosses zero at
radii unrelated to any resonance.

For each detected \rcr\ we also report an uncertainty interval.
Rather than imposing a fixed threshold, we take the interval to be
the contiguous range of radii about the zero-crossing over which
$|\Delta\phi|$ stays below a threshold that adapts to the local
steepness of the curve. Let $\Delta\phi_j$ denote the phase difference
at the $j$-th radius $R_j$ and
$\Delta\phi'_j = (\mathrm{d}\Delta\phi/\mathrm{d}R)_j$ its
finite-difference derivative. The threshold at ring $j$ is
\begin{equation}
T_j = \epsilon_0\left(1 + \frac{\alpha}{|\Delta\phi'_j|}\right),
\qquad
\epsilon_0 = \eta\left[\max_j \Delta\phi_j - \min_j \Delta\phi_j\right],
\label{eq:thresh}
\end{equation}
with $\eta = 0.01$ and $\alpha = 1\ \mathrm{rad}\,\rtw^{-1}$.
Beginning from the two rings that bracket the crossing, we extend the
interval outward in both directions for as long as
$|\Delta\phi_j| \le T_j$; the interval is the radial span of the rings
so accepted, floored at one ring spacing ($\pm\Delta R/2$ about the
candidate) when the walk terminates immediately. Because $T_j$ grows
as the curve flattens, a crossing that cuts steeply through zero is
bracketed tightly while a shallow one is assigned a wide interval.

\begin{figure}[t]
\centering
\includegraphics[width=0.95\textwidth]{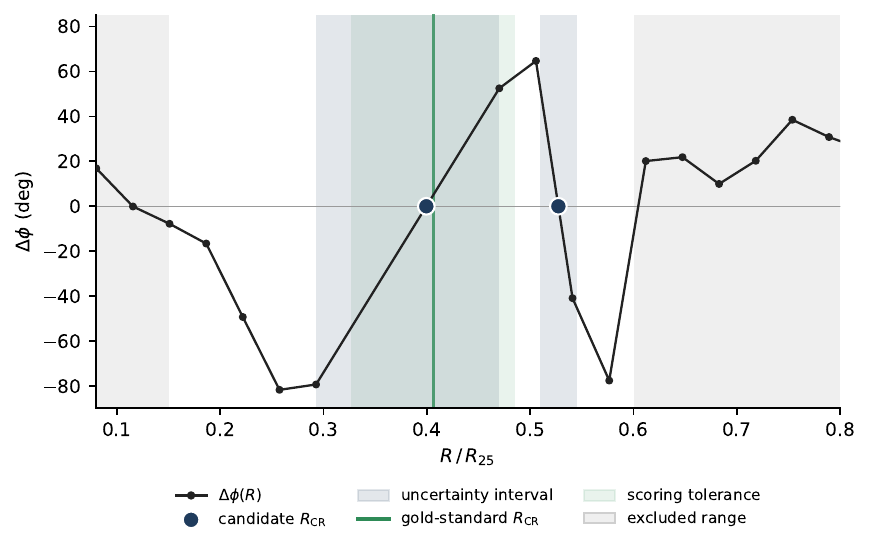}
\caption{Candidate corotation radii for NGC~6300.
The curve is the measured phase difference $\Delta\phi(R)$, which is
scanned for sign changes. Dark circles mark the two candidates, at
$0.40$ and $0.53\,\rtw$, with their uncertainty intervals shaded. The
vertical green line is the gold-standard value, $0.41\,\rtw$, and the
green band the tolerance used for scoring; the candidate at
$0.40\,\rtw$ coincides with it. Grey bands mark the radial range
excluded from candidate selection.}
\label{fig:methods:example}
\end{figure}


\section{Results}\label{sec:results}

\subsection{Reference set}\label{sec:results:validation}

We validate the pipeline against the compilation of \citet{Kostiuk2024},
which gathers published corotation radii measured with a wide range of
techniques. A meaningful cross-method test requires at least two independent
measurements of the same galaxy, so we restrict the compilation to the $161$
galaxies whose corotation radius has been measured by at least two different
methods, not including ours. These galaxies span a broad range of morphologies, sizes and distances,
and the published values for many of them disagree with one another, which makes
them a demanding benchmark for any new technique.

The pipeline was applied to all $161$ galaxies without any manual
intervention or galaxy-specific tuning. Of these, $93$ lie in the SDSS footprint and
were analysed in the $g$ and $i$ bands, and $68$ were analysed with DSS2 blue and
WISE $3.4\,\mu$m images (Section~\ref{sec:methods:profile}). The pipeline yields a
measurement for $161$ galaxies.

The reference against which every method is scored is built from the published
measurements alone. For each galaxy we intersect the reported corotation intervals
of every pair of distinct literature sources and merge the non-empty intersections
into a disjoint set; a galaxy enters this gold standard when at least two
independent sources agree. Requiring agreement between independent sources
suppresses the method-specific scatter that dominates individual measurements,
and leaves a reference that no single technique controls. Our own measurements
take no part in its construction, and each published method is scored against a
gold standard rebuilt without its own measurements, so that no method is compared
with a reference it helped to define.

\subsection{Scoring and published methods}\label{sec:results:scoring}

The resulting gold standard contains 90 galaxies, $58$ from SDSS imaging and $32$ from the all-sky surveys. A predicted
corotation radius is counted as correct when it lies within
$0.10\,\rtw$ of the midpoint of a gold-standard interval, where \rtw\ is the
isophotal radius quoted in the compilation, so that every method is measured on
the same scale. The criterion depends only on the location of each predicted
radius, not on the width of its reported uncertainty, and therefore applies in the
same way to methods that quote intervals and to those that quote single values.
Published methods are scored by their reported \rcr\ values.

From this criterion we compute five quantities for each method. The
\emph{completeness} is the fraction of scored galaxies with at least one correct
\rcr; it measures how often a method recovers the corotation radius of a galaxy.
The \emph{recall} is the fraction of all gold-standard corotation radii that are
matched by at least one predicted radius; it measures how many of the known
resonances a method finds. The \emph{purity} is the fraction of all predicted radii
that are correct; it measures on average how reliable an individual prediction is. The
harmonic mean of recall and purity, the $F_1$ score, combines them into a single
figure of merit that is high only when a method finds most of the known resonances
and is reliable in doing so. Finally, the \emph{accuracy per
galaxy} is the fraction of a galaxy's predicted radii that are correct, averaged
over galaxies, which weights every galaxy equally regardless of how many radii a
method reports for it.

We compare the pipeline with every published method in the compilation that
can be scored on at least $16$ galaxies. Six methods meet this requirement, and
together they represent the main families of corotation diagnostics: the
potential--density phase-shift method of \citet{Buta2009}, applied to
near-infrared images; the multi-wavelength azimuthal offset method of
\citet{Abdeen2020}; the bar-torque analysis of \citet{RuizGarcia2024}; the
morphological criteria of \citet{Elmegreen1995}; the kinematic
Tremaine--Weinberg measurements of \citet{Williams2021}; and the numerical
models of \citet{Rautiainen2008}. All of these were applied to their
samples individually, with substantial human input, whereas our pipeline runs
unsupervised. The results are summarised in Table~\ref{tab:results:summary}.

\begin{table}[htbp]
\caption{Agreement of each method with the literature gold standard (criterion:
predicted \rcr\ within $0.10\,\rtw$ of a gold-standard value). $N$ is the number
of galaxies scored, $N_{\rm c}$ the number with a correctly located \rcr, and
$F_1$ the harmonic mean of recall and purity. Recall and purity are counted per
corotation radius, completeness per galaxy; all three are conditional on the
galaxies each method reports on, and $N$ therefore differs between rows.}
\label{tab:results:summary}
\centering
\footnotesize
\setlength{\tabcolsep}{3pt}
\begin{tabularx}{\textwidth}{@{}>{\raggedright\arraybackslash}X
                               >{\raggedright\arraybackslash}X
                               rrrrrr@{}}
\toprule
Method & Type & $N$ & $N_{\rm c}$ & Compl. & Recall & Purity & $F_1$ \\
\midrule
This work & phase offset & 90 & 72 & \textbf{80.0}\% & \textbf{63.6}\% & 52.5\% & \textbf{0.58} \\
\citet{Buta2009} & potential--density & 40 & 29 & 72.5\% & 57.7\% & 35.9\% & 0.44 \\
\citet{Abdeen2020} & multi-band offset & 16 & 11 & 68.8\% & 52.5\% & \textbf{56.2}\% & 0.54 \\
\citet{RuizGarcia2024} & bar torque & 21 & 11 & 52.4\% & 38.5\% & 52.4\% & 0.44 \\
\citet{Elmegreen1995} & morphological & 44 & 22 & 50.0\% & 29.4\% & 51.1\% & 0.37 \\
\citet{Williams2021} & Tremaine--Weinberg & 19 & 9 & 47.4\% & 40.5\% & 47.5\% & 0.44 \\
\citet{Rautiainen2008} & models & 19 & 9 & 47.4\% & 25.7\% & 47.4\% & 0.33 \\
\bottomrule
\end{tabularx}
\end{table}

\subsection{Galaxy-level agreement}\label{sec:results:compare}

\begin{figure}[t]
\centering
\includegraphics[width=0.9\textwidth]{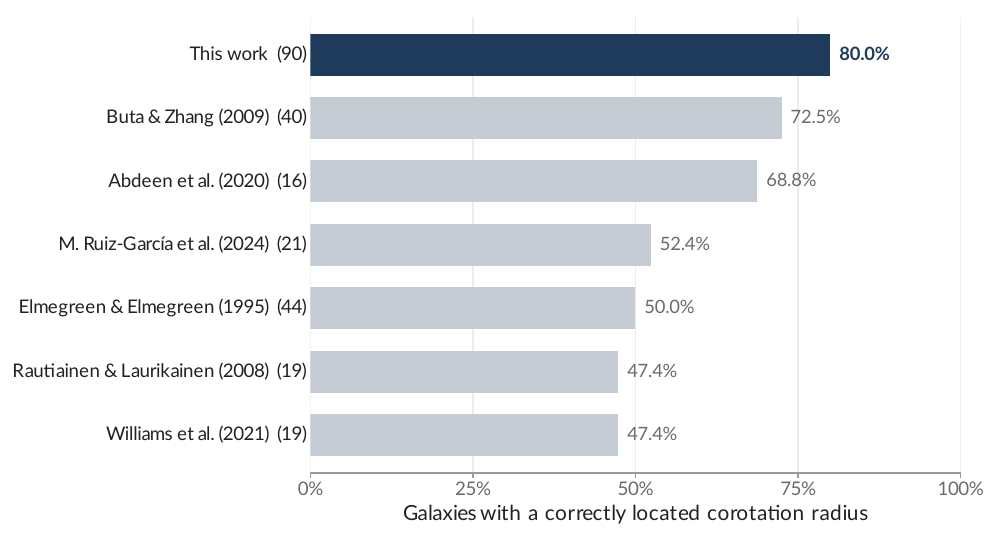}
\caption{Fraction of galaxies in which at least one predicted \rcr\ lies within
$0.10\,\rtw$ of the gold standard. Numbers in brackets give the number of galaxies
scored. Each published method is scored against a gold standard built without its
own measurements.}
\label{fig:results:galaxy}
\end{figure}

Our pipeline places a correct corotation radius in $72$ of its $90$ scored
galaxies, a galaxy-level agreement of $80.0\%$, the highest of all the methods
considered (Fig.~\ref{fig:results:galaxy}). The next highest are \citet{Buta2009},
with $72.5\%$ ($29$ of $40$ galaxies), and \citet{Abdeen2020}, with
$68.8\%$ ($11$ of $16$). The remaining four methods, including the kinematic
Tremaine--Weinberg measurements and the dynamical models, recover the corotation
radius in only $47$--$52\%$ of their galaxies.

This result is notable because our pipeline obtains it without supervision and
from photometry alone. The methods it outperforms include techniques that require
integral-field kinematics or detailed dynamical modelling, and that were applied by
hand to carefully selected galaxies. Our pipeline reaches a higher agreement on a
sample that was not selected for the quality of its spiral structure and that
includes images from several different surveys.

\subsection{Recall and purity}\label{sec:results:purity}

\begin{figure}[t]
\centering
\includegraphics[width=0.9\textwidth]{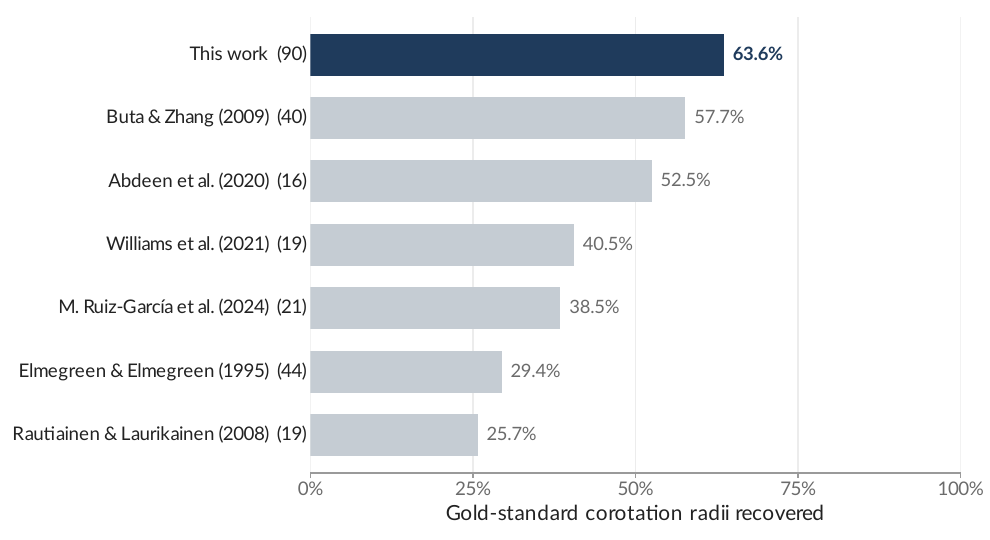}
\caption{Fraction of gold-standard corotation radii recovered by at least one
predicted \rcr\ within $0.10\,\rtw$ (recall). Numbers in brackets give the number of
galaxies scored.}
\label{fig:results:recall}
\end{figure}

Galaxies frequently host more than one corotation radius, and the gold
standard contains $195$ corotation radii for the $90$ galaxies scored. Our pipeline
recovers $124$ of them, a recall of $63.6\%$, the highest of the methods compared
(Fig.~\ref{fig:results:recall}). It is followed by \citet{Buta2009}, with
$57.7\%$, and \citet{Abdeen2020}, with $52.5\%$; the remaining methods recover
between $26\%$ and $41\%$ of the known corotation radii in their galaxies.

\begin{figure}[t]
\centering
\includegraphics[width=0.9\textwidth]{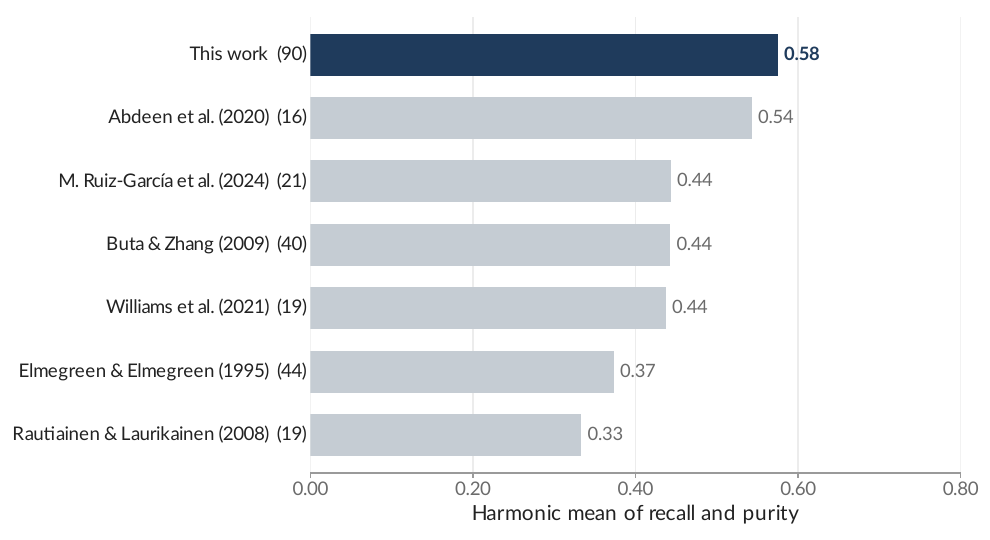}
\caption{Harmonic mean of recall and purity (the $F_1$ score), for the same
methods and criterion as Fig.~\ref{fig:results:galaxy}.}
\label{fig:results:harmonic}
\end{figure}

A method can achieve high recall by reporting many radii for each galaxy, at
the cost of reliability, or high purity by reporting only its most secure radius, at
the cost of recall. The harmonic mean of the two rewards methods that achieve both.
Our pipeline reaches $F_1 = 0.58$, from a recall of $63.6\%$ and a purity of
$52.5\%$ (Fig.~\ref{fig:results:harmonic}), again the highest value among the
methods compared. It is followed by \citet{Abdeen2020} at $0.54$;
\citet{Buta2009}, \citet{RuizGarcia2024} and \citet{Williams2021} reach
$0.44$, and the remaining two methods $0.33$--$0.37$.

The comparison with \citet{Buta2009} illustrates why both quantities are
needed. Their galaxy-level agreement, $72.5\%$, and their recall, $57.7\%$, are the
second highest, but only $35.9\%$ of their individual radii are correct, giving
$F_1 = 0.44$.
The pipeline's purity of $52.5\%$ is among the highest in the comparison, close to
that of \citeauthor{Abdeen2020} (\citeyear{Abdeen2020}; $56.2\%$) and comparable to those of
\citeauthor{RuizGarcia2024} (\citeyear{RuizGarcia2024}; $52.4\%$) and
\citeauthor{Elmegreen1995} (\citeyear{Elmegreen1995}; $51.1\%$), so its high completeness is not obtained at the expense of reliability.

\subsection{Accuracy and coverage}\label{sec:results:coverage}

\begin{figure}[t]
\centering
\includegraphics[width=0.9\textwidth]{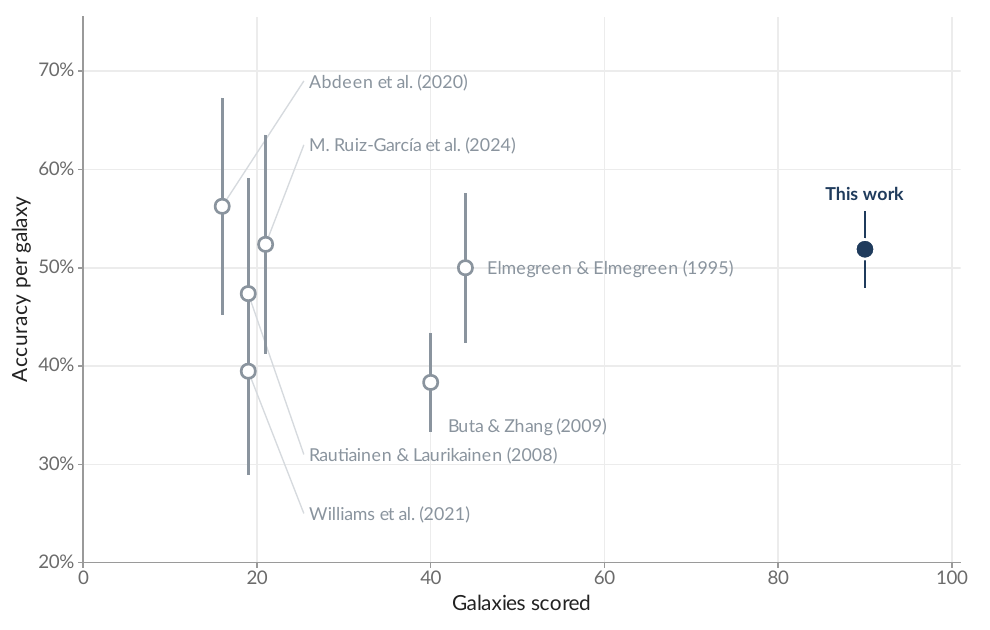}
\caption{Accuracy per galaxy, the mean fraction of a galaxy's predicted radii
within $0.10\,\rtw$ of the gold standard, against the number of galaxies scored.
Error bars are one standard error over galaxies.}
\label{fig:results:coverage}
\end{figure}

Figure~\ref{fig:results:coverage} places the accuracy of each method against the
number of galaxies on which it can be scored. Our pipeline's accuracy per galaxy,
$51.9 \pm 3.9\%$, lies within the range of the published methods ($38$--$56\%$),
and its uncertainty is the smallest in the comparison, because its scored sample of
$90$ galaxies is two to six times larger than theirs ($16$--$44$ galaxies). No
published method combines a comparable accuracy with a comparable sample size.

In absolute terms, our pipeline recovers the corotation radius in $72$ galaxies,
more than twice as many as any single published method, the largest of which
recovers $29$. Because the pipeline requires only two-band imaging and no manual
input, this number is limited by the size of the reference set rather than by the
method itself, and it can be applied directly to the much larger samples provided by
current and forthcoming imaging surveys.

\section{Discussion and conclusions}\label{sec:discussion}

Although the spiral mode number and the fitting function are configurable, in this work we fix $m=2$ and use a single cosine model. While a cosine model captures most symmetric structures
effectively, alternative periodic or data-driven functions can be
substituted to model higher-order, asymmetric, or transient patterns.
This flexibility allows the method to adapt to a wide range of
spiral morphologies while maintaining quantitative consistency.

Where $\Delta\phi(R)$ shows a clear reversal of sign, as in
NGC~6300 (Fig.~\ref{fig:methods:example}), the derived \rcr\ coincides
with the published value. In systems showing multiple
resonances, our method detects several distinct zero-crossings,
consistent with coupled or evolving patterns reported previously.
These results support a hybrid interpretation of spiral structure in
which both quasi-stationary density waves and transient, bar-driven,
or tidally induced features can coexist within the same disk.

The principal limitations of the method arise in galaxies with weak
or irregular arm structure, where stochastic star-forming regions
dominate the brightness signal. Restricting candidates to the radial range of
Section~\ref{sec:methods:uncertainty} removes most spurious crossings,
but such morphologies remain challenging. Incorporating multi-band phase
analysis, mode-adaptive fitting, and improved noise filtering may
further enhance robustness for these cases.

Validated against a
cross-method reference catalog, our approach achieves the highest galaxy-level agreement, the highest recall and the highest harmonic mean of recall and purity among the methods compared, while operating without supervision on a larger sample. The published
techniques were applied with substantial manual work to samples selected by their
authors, and several quote wide uncertainty intervals (a median of $31''$ for
\citealt{Abdeen2020} and $34''$ for \citealt{Williams2021}), whereas the pipeline runs on every
galaxy of the reference set with intervals a median $7.5''$ wide. Its
adaptability, both in the chosen spiral mode number and in the
fitting function, extends applicability to grand-design, multi-arm,
and flocculent spirals alike. The observed diversity of single and
multiple resonances across our sample provides new large-scale
evidence that spiral galaxies may host a combination of stable and
transient dynamical modes.

Future work will focus on testing the method with alternative
parameters and functional models to further assess its stability and
precision across different galaxy types. We plan to study correlations
between the detected corotation radii and known galactic features,
such as bar presence and morphology. Finally, we aim to build an
easy-to-use tool that allows astronomers to apply this method to
their own datasets and to publish a public database of galaxies with
automatically detected corotation radii. Together, these efforts will
make the automated photometric color-gradient framework a practical and
widely accessible resource for future studies of galactic dynamics.


\backmatter






\bibliography{sn-bibliography}

\end{document}